\documentclass[reprint,amsmath,amssymb,aps]{revtex4-2}

\usepackage{graphicx}% Include figure files
\usepackage{dcolumn}% Align table columns on decimal point
\usepackage{bm}% bold math
\usepackage{tabularx,multirow,array,booktabs,makecell,float}
\usepackage[utf8]{inputenc}
\usepackage[T1]{fontenc}

\begin{document}

%\preprint{APS/123-QED}

\title{GHz-Pulsed Source of Entangled Photons for Reconfigurable Quantum Networks}

\author{Meritxell Cabrejo Ponce}
\email{meritxell.cabrejo.ponce@iof.fraunhofer.de}

\author{Christopher Spiess}%
\affiliation{Friedrich Schiller University, Max-Wien-Platz 1, 07743 Jena, Germany
}%
\affiliation{Fraunhofer Institute for Applied Optics and Precision Engineering, Albert-Einstein-str. 7, 07745 Jena, Germany}

\author{André Luiz Marques Muniz and Philippe Ancsin}
\affiliation{Fraunhofer Institute for Applied Optics and Precision Engineering, Albert-Einstein-str. 7, 07745 Jena, Germany
}%

\author{Fabian Steinlechner}
\email{fabian.steinlechner@iof.fraunhofer.de}
\affiliation{Fraunhofer Institute for Applied Optics and Precision Engineering, Albert-Einstein-str. 7, 07745 Jena, Germany
}%
\affiliation{Abbe Center of Photonics, Friedrich Schiller University Jena, Albert-Einstein-Str. 6, 07745 Jena, Germany}

\date{\today}% It is always \today, today,

\begin{abstract}
Entanglement is a universal resource in quantum networks, yet entangled photon sources are typically custom-made for a specific use case.  Versatility, both in terms of state modulation and tunability of the temporal properties of the photons, is the key to flexible network architectures and cryptographic primitives that go beyond quantum key distribution. Here, we report on a flexible source design that produces high-quality entanglement in continuous-wave and GHz-rate-pulsed operation modes.
Utilizing off-the-shelf optical components, our approach uses a fiber-based Sagnac loop to generate polarization-entangled photons at telecom wavelength with high efficiency and fidelities above 0.99. Phase modulation up to GHz before entangled state generation is also possible for fast entangled state switching. We show phase modulation at 100 MHz with an average fidelity of 0.95.
Furthermore, the source 60 nm spectral bandwidth is entirely compatible with fully reconfigurable wavelength-multiplexed quantum networks.\end{abstract}


\maketitle

%%%%%%%%%%%%%%%%%%%%%%%%%%  body  %%%%%%%%%%%%%%%%%%%%%%%%%%
\section{Introduction\\}
Entangled photons are an essential resource in emerging quantum information processing and secure quantum communications \cite{steane_quantum_1998, slussarenko_photonic_2019}. In the context of quantum cryptography, entanglement-based quantum key distribution (QKD) has recently received considerable attention due to its inherent simplicity. 
It leverages the intrinsic randomness of quantum entanglement and can be implemented with passive optical components.
Moreover, it allows for scalable multiuser operation through the use of auxiliary frequency correlations for multiplexing \cite{wengerowsky_entanglement-based_2018, joshi_trusted_2020}.
However, richer cryptographic primitives and beyond require more flexibility from the source. 
Desiderata for protocols like quantum teleportation and entanglement swapping are tunability from continuous-wave to pulsed operation. On the other hand, fast state control is also required for reconfigurable quantum networks \cite{williams_quantum_2019}
and particularly for protocols such as three-partite quantum secret sharing (QSS), that need phase modulated entangled states \cite{grice_reconfigurable_2019}. 

Typical sources are those of polarization entangled qubits, often consisting of free-space interferometers based on bulk optics \cite{anwar_entangled_2021-1,brassard_limitations_2000, lee_sagnac-type_2021}. They provide stable operation and high photon-pair rates by employing spontaneous parametric down-conversion (SPDC) processes in $\chi^{(2)}$-nonlinear crystals or waveguides \cite{levine_polarization-entangled_2011, fiorentino_source_2005, martin_polarization_2010, caspani_integrated_2017}. However, the dimensions of the bulky components make it difficult for the source to be reconfigurable and especially fast. They are optimized to be solely used for continuous wave or in pulsed regimes and are not easily interchangeable. Furthermore, pump and down-converted photons frequently need to propagate along the same paths, involving the use of expensive dual-wavelength components to achieve high visibility  \cite{lee_sagnac-type_2021, kim_pulsed_2019, li_cw-pumped_2015, vergyris_fully_2017}. 

A lot of effort has been directed toward the realisation of polarization entangled photon sources (EPS) with both pump and correlated photons in telecom wavelengths. In this manner, inexpensive high-speed optical modulators and communication through the already existing global fiber network could be exploited. 
The natural choices for those systems are $\chi^{(3)}$ materials via spontaneous four wave mixing, since they enable both pump and photon-pairs in the same wavelength bands \cite{suo_generation_2015, zhang_generation_2019}. Nevertheless, their bulk nonlinearity is $10^4-10^5$ orders of magnitude smaller than cascaded $\chi^{(2)}$ processes (i.e. second harmonic generation and SPDC in the same $\chi^{(2)}$ material) \cite{chou_15-m-band_1999,jiang_generation_2007,arahira_generation_2011}.

\begin{figure*}[t!]
\centering\includegraphics[width=0.8\textwidth]{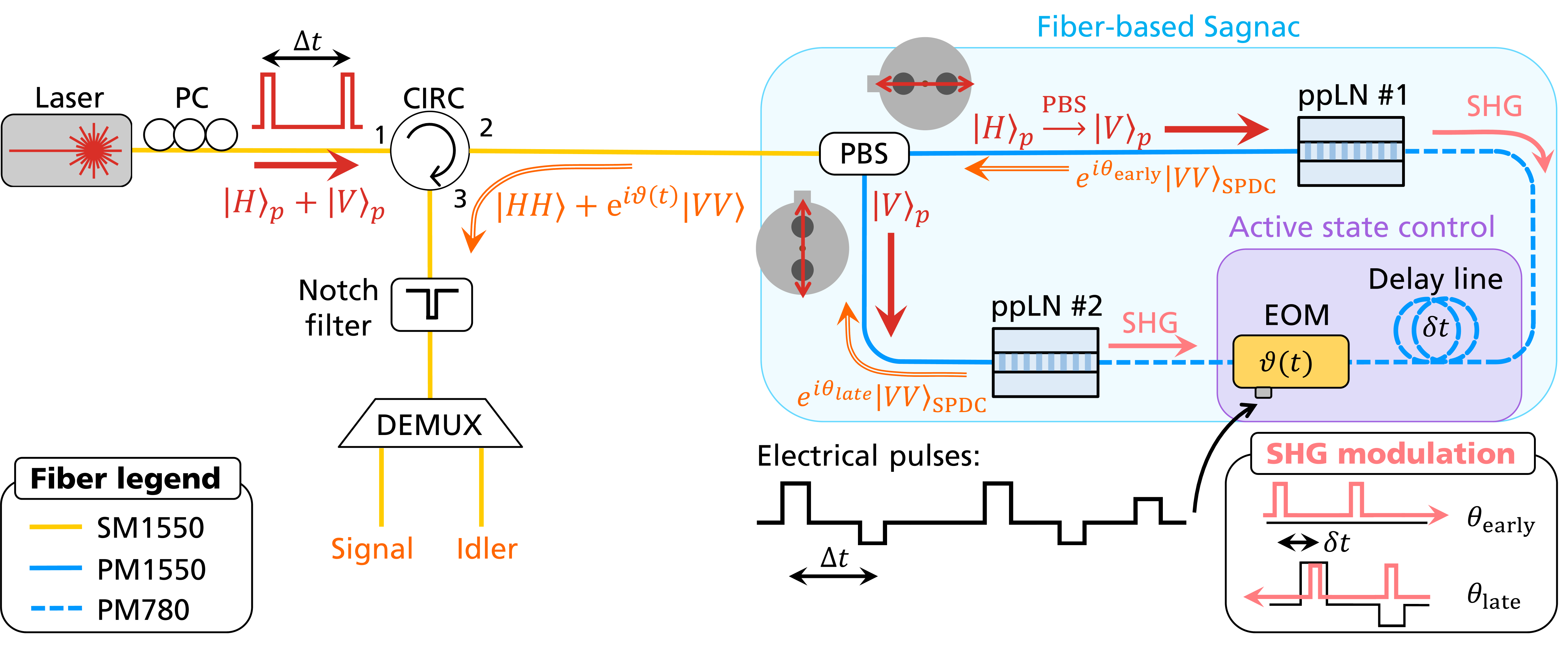}
\caption{Schematic of our polarization entangled photon-pair source: a telecom pump (continous-wave or pulsed) is sent to the fiber-based Sagnac interferometer (blue box). It consists of a fiber loop with bidirectional up- and down-conversion stages. After second harmonic generation (SHG), phase modulation $\vartheta(t)=\theta_\text{early}(t)-\theta_\text{late}(t)$ can be implemented (purple box) with arbitrary electrical signals and same period $\Delta t$ as the optical pulses. In that case, a delay $\delta t$ is employed to temporally separate and independently address the SHG pulses. The emerging state is rerouted with a circulator (CIRC) to our detection system. Single-mode (SM) and polarization-maintaining (PM) fibers are shown in yellow and blue color, respectively. EOM: electro-optic modulator, PBS: polarizing beam splitter, PC: polarization controller, ppLN: periodically poled Lithium Niobate.}
\label{fig:Setup}
\end{figure*}

In this letter, inspired by the work in \cite{agnesi_all-fiber_2019, li_experimental_2019, li_high-speed_2019, alibart_quantum_2016, tanzilli_genesis_2012-1} for faint-pulse sources (i.e., a classical laser attenuated to the single photon level), we propose a novel EPS at telecom wavelengths based on a fiber-based Sagnac interferometer with two $\chi^{(2)}$ nonlinear waveguides. 
In this configuration, a telecom pump enters the fiber loop and follows a cascaded up- and down-conversion process in both directions. The immediate advantage is that the source does not need dual-wavelength components like the free-space versions. 
Additionally, the design allows the incorporation of fiber-integrated optical modulators to i) easily switch the pump from continuous-wave (CW) to GHz-pulsed regimes, and ii) to modulate the phase of the generated state up to GHz rate. 
To demonstrate the versatility of this source, we characterize it in CW and GHz-pulsed regimes in terms of photon pair generation and visibility as a function of repetition rate and duty cycle for coarse and dense wavelength division multiplexing (WDM) arrangements. 
We report Bell-state fidelities above 0.99 for different pump configurations and high brightness, with millions of photons at emission per nm of bandwidth. The spectrum spans over 60 nm, offering the possibility to increase the number of users via wavelength demultiplexing or to actively allocate the necessary bandwidth in a practical scenario \cite{appas_flexible_2021}.

%Corresponding to 350-fold higher actual pair-rate generation, our system considerably exceeds state-of-the-art pulsed sources \cite{slussarenko_photonic_2019, kim_pulsed_2019}. 
Finally, to showcase the ultimate flexibility of our design, we continuously switch the generated entangled state from a basis of four different states. In this proof-of-concept experiment, we achieve average state fidelities of 0.95 at 100 MHz rate. These results are a step forward towards a fully reconfigurable entangled photon source and may pave the way for active quantum networks.

\section{Entangled photon source}
\begin{figure}
\centering
\includegraphics[width=0.4\textwidth]{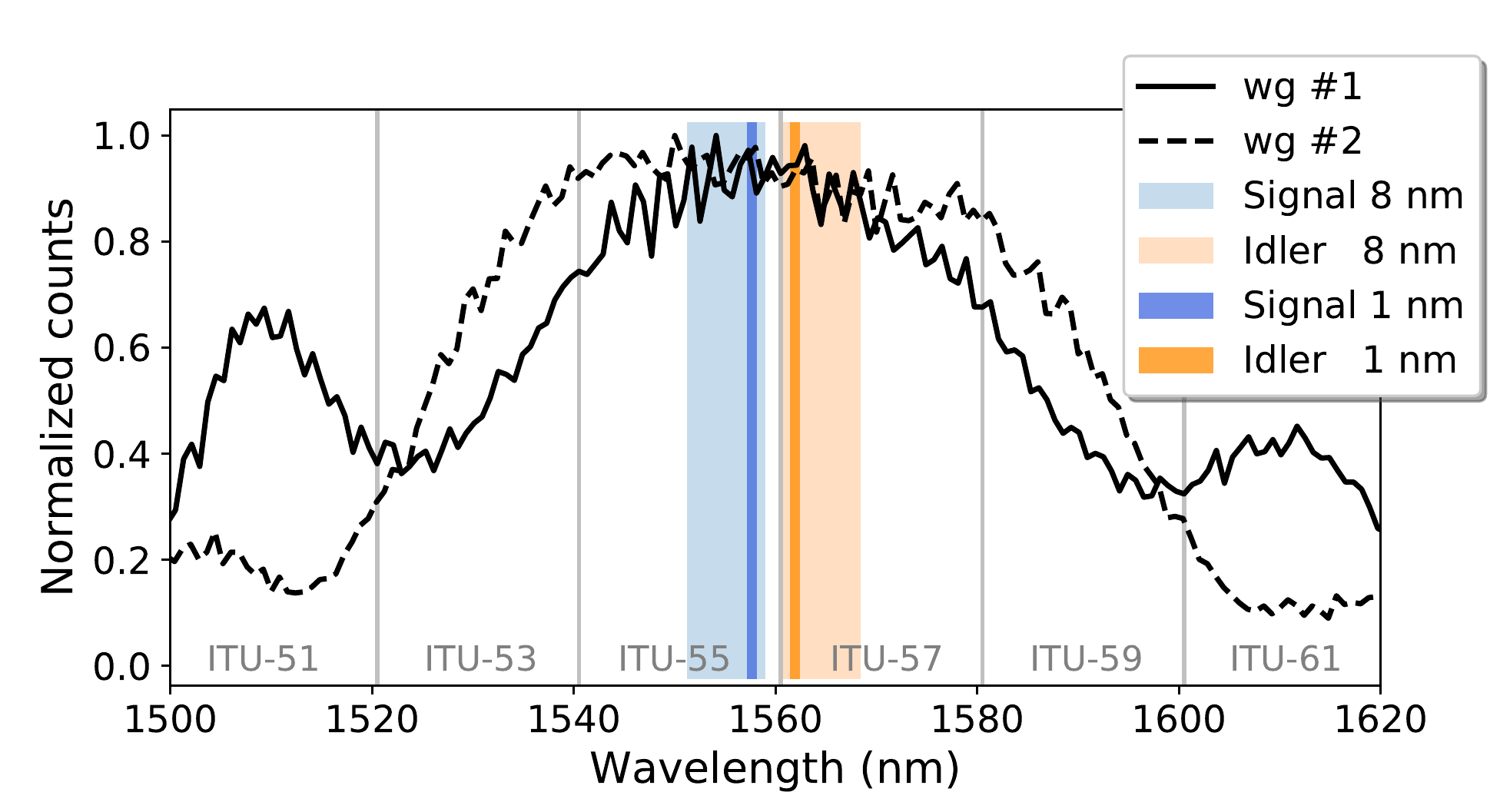}
\caption{SPDC spectra of ppLN waveguides (wg) \#1 and \#2. The coloured areas indicate the channel bandwidth of the signal and idler for our coarse (8 nm) and dense demultiplexing (1 nm) settings. Yet, our source is also compatible with the international telecom union (ITU) standard coarse wavelength division multiplexing channels (CWDM, 20 nm grid).}
\label{fig:Spectra}
\end{figure}

Our fiber-based Sagnac loop is pictured in Fig. \ref{fig:Setup}. After pump preparation (see Methods), the pump polarization is set with polarization controllers (PC) right before a polarization beam splitter (PBS). 
By launching 45$^{\circ}$-linearly polarized optical pulses, $|\Psi_p\rangle=\frac{1}{\sqrt{2}}(|H\rangle_p+|V\rangle_p)$, the fiber PBS accordingly separates H-and V-components and sends them over polarization-maintaining (PM) fibers, %in clockwise and counterclockwise directions
whereby each output is aligned to the slow axis of the fiber.
Thus, both H and V polarizations travel along the same axis such that the group velocity mismatch is compensated. The outputs of the PBS are connected to two fiber-pigtailed periodically poled lithium niobate (ppLN) waveguides (Covesion MSHG1550 with 40mm long type-0 crystal). 
These are commercially available second harmonic generation (SHG) modules, where a 1560 nm PM fiber is directly coupled to a ppLN waveguide, while a 780 nm PM fiber is coupled at the other end. They have slightly different SHG conversion efficiencies due to fabrication imperfections, $10\%$ and $11\%$ respectively for 300 mW of pump power. 
Optimum SHG conversion is achieved by tuning the temperature of the ppLN waveguides \#1 and \#2 (53°C and 48°C respectively), which simultaneously produces broadband pair generation around a degeneracy wavelength of 1560 nm. 
The SPDC spectrum spans over the C-band and beyond (from 1530 to 1590nm) with the full width at half maximum (FWHM) of 58 nm and 68 nm for waveguide \#1 and \#2 respectively, as shown in Fig. \ref{fig:Spectra}. 
Finally, both ppLN modules are interconnected with a 780 nm PM fiber to close the loop. 
As a result, the second harmonic generated light of one waveguide is sent to the opposite one to generate SPDC and vice versa. 
This situation occurs %simultaneously 
in both clockwise and counterclockwise directions, giving rise to a polarization-entangled state. 
Additionally, any perturbance on the fiber loop affects simultaneously both propagation directions, maximizing phase stability of the entangled state.
For active state switching, as explained at the end of this letter, we can introduce here a delay line to temporally separate the SHG pulses, and an electro-optic modulator (EOM) to modulate their relative phase $\vartheta(t)$ (purple box in Fig. \ref{fig:Setup}). 

In this configuration, SPDC photon pairs from both ppLN modules are combined at the output of the PBS and emerge in the entangled state $|\phi\rangle = \alpha |H_s H_i \rangle + e^{i\vartheta(t)} \beta |V_s V_i \rangle$, where $H_{s(i)}$ is the signal (idler) photon in the $H$ polarization, and similarly for $V$. In the remaining, we suppress the subscripts of the state.
Here, $\alpha$ and $\beta$ are the probabilities of generating SPDC photon-pairs clockwise and counterclockwise, respectively, while $\vartheta(t)$ denotes the detuning phase between both directions. 
For a 45° linearly polarized pump and without SHG modulation, $\alpha = \beta = 1/\sqrt{2}$ and $\vartheta=0$, resulting in the maximally entangled Bell state $|\phi^+\rangle$.
The outgoing light from the PBS is rerouted via the circulator (CIRC) followed by a 100-dB notch filter, with 1.6 nm bandwidth at 1560 nm, that removes any remaining fraction of the pump or additional noise. 
Finally, the signal and idler correlated photons are separated by a demultiplexer (DEMUX) to be sent to independent users. Here, we directly send them to our detection system (see Methods).

\section{Performance analysis}
\begin{figure}[ht]
	\centering
\includegraphics[width=0.49\textwidth]{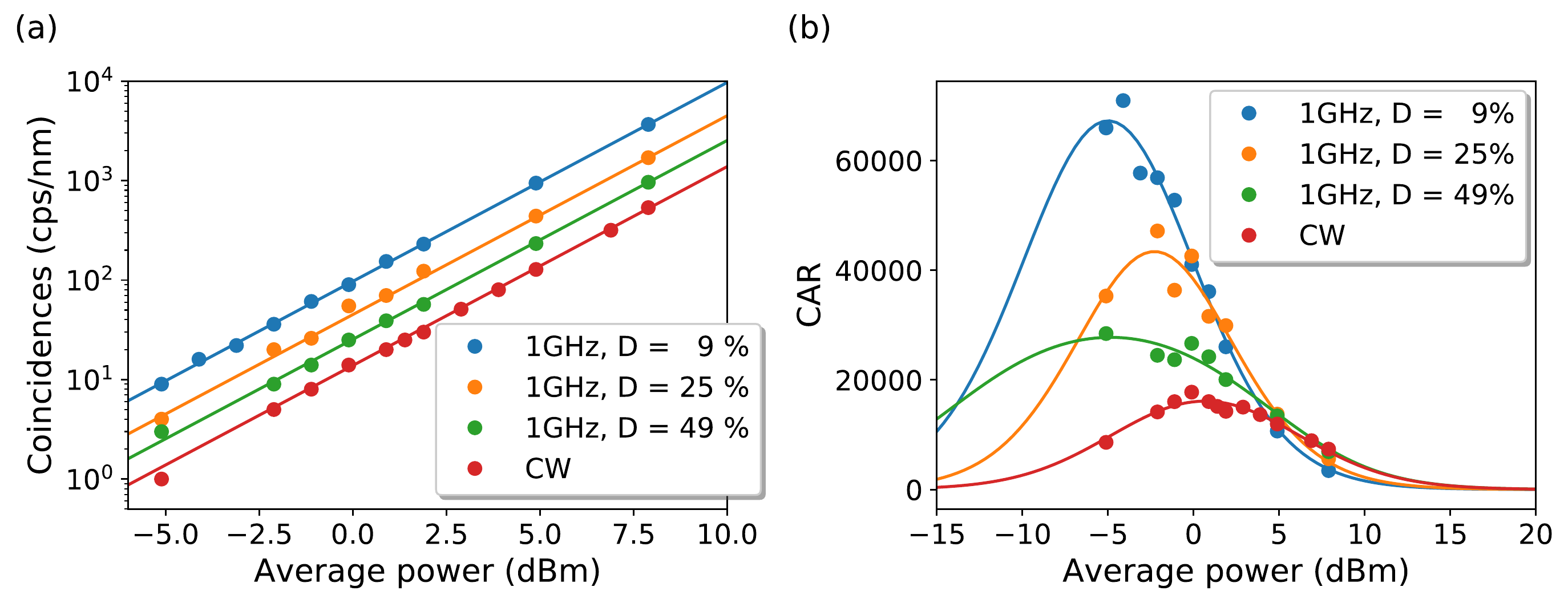}
\caption[Performance]{Analysis of the source performance under CW and pulsed regimes at 1 GHz and various duty cycles: (a) Coincidence counts per nm of spectral bandwidth as a function of average pump power at 1560 nm, where the lines correspond to a power of 2 fit. The photon pair rate increases quadratically with pump power due to the SHG process. Moreover, shorter duty cycles also enhance the SHG conversion per pulse, resulting in greater coincidence events. (b) CAR results for the same system settings, where the lines correspond to our fitted model including noise. Notice that lower duty cycles shift the peak towards lower powers, but also increase CAR values thanks to improved SHG and lower noise counts. }
    \label{fig:Performance}
\end{figure}

We first characterize our source excluding the active state control elements. In our narrowband WDM scheme, the signal and idler channels selected by the programmable filter are separated by 3.2 nm with a bandwidth of $\Delta\lambda=$ 1 nm, as shown in Fig. \ref{fig:Spectra}. 
We report in Fig. \ref{fig:Performance}a the photon-pair coincidence counts per nm of spectral bandwidth, a parameter known as the spectral brightness, as a function of the average pump power ($P_\text{in}$) right before the PBS, for CW and pulsed operation at repetition rates of 1 GHz and duty cycles of 9\%, 25\% and 49\%. 
These measurements are performed over either H- or V-polarized signal and idler SPDC photons, which represent the performance of a single direction of the loop. 
The coincidence counts grow with a power of 2 and are proportional to $P_\text{in}^2$ (fitted in Fig. \ref{fig:Performance}a), according to the SHG power conversion. 
Furthermore, shorter duty cycles yield stronger SHG at the same average but higher peak pump power, and therefore, higher SPDC brightness.
The photon-pair heralding, defined as the ratio of coincidences over single-photon counts, improves with pump power and shorter duty cycle because of a relative reduction of additional noise sources (see Supplementary Material). 
It amounts to $H_s=1.6-2.9\%$ and $H_i=2.9-5.2\%$ for signal and idler, respectively, differences that can be explained by the accumulated and asymmetric losses in the system as well as the coupling efficiency of the ppLN waveguide modules. 
With a total loss of $\sim$ 12.85 dB (see Methods), the actually emitted heralding is almost 20-fold higher. 

To verify the quality of the entanglement, polarization projecting measurements are performed in two mutually unbiased bases, H/V and D/A, for different pump parameters. In the H/V basis and for the state $|\phi^+ \rangle$, the visibility per nm of spectral bandwidth is computed as: 

\begin{equation}
V_{HV} =
 \frac{R_{HH} - R_{VH} - R_{HV} + R_{VV}}
 {R_{HH} + R_{VH} + R_{HV} + R_{VV}}
\end{equation}
where $R_{ij}$ denotes the coincidence rate measured with i, j settings, and similarly for D/A. We find that for a 1 GHz repetition rate and $9\%$ duty cycle, we achieve visibilities of $99.5\%$ in H/V and $99.0 \%$ in D/A, with no subtraction of accidentals. Using the fidelity witness $F\geq(V_{HV}+V_{DA})/2$  \cite{anwar_entangled_2021}, we lower-bound the Bell-state fidelity to $F=99.25\%$.

\begin{table*}[t!]
    \caption{Detected and estimated photon pair rates ($R^\text{exp}$ and $R^\text{est}$) and heralding values for signal and idler photons ($H_s^\text{exp}$, $H_i^\text{exp}$ and $H_s^\text{est}$, $H_i^\text{est}$) at the output of the entangled photon source (i.e. after the PBS) for one polarization projection $|\Psi_\text{proj}\rangle=|HH\rangle$. Notice that the heralding is limited by the coupling of the fibers to the ppLN waveguides. $P_\text{in}$: input pump power right before the PBS, $\Delta\lambda$: signal and idler channel bandwidth. }
	\label{table:source_data}
  \centering
  \small
    \begin{tabular}{c >{\centering}m{2.4cm} c c|c c c|c c c} \toprule
        %\multirow{2}{2.5cm}{\centering Pump rep. rate} & 
        %\multirow{2}{1.5cm}{\centering Duty cycle (\%)} &
        Pump rep. rate &
        Duty cycle (\%) &
        $P_\text{in}$ (dBm) & 
        $\Delta\lambda$ (nm) & 
        $R^\text{exp}$ (cps)  & 
        $H_s^\text{exp}$ (\%) &
        $H_i^\text{exp}$ (\%) &
        $R^\text{est}$ (Mcps)  & 
        $H_s^\text{est}$ (\%) &
        $H_i^\text{est}$ (\%) \\
        %& & & & & & & & & \\
        \midrule
        CW                      & 100 & 7.9 & 1 &  535 & 1.6 & 2.9 & 0.20 & 39.8 & 44.2 \\ \cmidrule{1-10}
        \multirow{2.3}{2.4cm}{\centering 1 GHz} &  9 & 7.9 & 1 & 3672 & 2.9 & 5.2 & 1.36 & 71.6 & 79.0 \\ \cmidrule(l){2-10}
         & 49 & 7.9 & 1 &  963 & 2.1 & 3.8 & 0.36 & 51.4 & 57.6 \\ \cmidrule{1-10}
         4 GHz                    & 44 & 7.9 & 1 &  901 & 2.1 & 3.8 & 0.33 & 51.7 & 57.3 \\ \cmidrule{1-10}
        \multirow{2.3}{*}{8 GHz} & 50 & 7.9 & 1 &  565 & 1.9 & 3.3 & 0.21 & 45.5 & 50.0 \\ \cmidrule(l){2-10}
         & 50 & 9.0 & 8 & 8027 & 1.7 & 3.6 & 2.98 & 41.6 & 54.0 \\ \bottomrule
    \end{tabular}
    \normalsize 
\end{table*}

Next, we assess the contribution of additional noise sources to the experimentally observed heralding efficiency. We evaluate the coincidence-to-accidental ratio (CAR) for CW and pulses at 1 GHz repetition rates with the same duty cycles as before and display the results in Fig. \ref{fig:Performance}b. The CAR parameter is defined by the ratio between true photon-pair coincidences generated within the same SPDC process and accidental coincidences with subsequent pairs and noise photons. The accidental detections are proportional to the single-photon counts of each channel as well as the total coincidence window. Accordingly, we can model the CAR as a function of input power as:

\begin{equation}
CAR =
 \frac{\mu \eta_s \eta_i}
 { ( (\mu + N_{noise})\eta_s + d_s ) ( (\mu + N_{noise})\eta_i + d_i ) } - 1
\end{equation}
where $\mu$ denotes the average photon-pair number per coincidence window, $\eta_{s(i)}$ stand for the detection efficiencies and $d_{s(i)}$ are the dark count probabilities for the signal (idler) channels. 
$N_{noise}$ is an additional noise term accounting for excess single-photon contributions attributed to spontaneous Raman scattering (SpRS), further discussed in the Supplementary Material. 
Detection efficiency, dark counts ($\sim$300 Hz), and noise photon numbers are expected to be similar for both channels. 
The relation of SPDC and SpRS noise with pump power $P_\text{in}$ follows as $\mu = \gamma_\text{SPDC} \gamma_\text{SHG} P_\text{in}^2$ and $N_\text{noise}=\gamma_\text{noise} P_\text{in}$ \cite{eraerds_quantum_2010}, with $\gamma$ including the respective generation efficiencies. 
Given the large number of parameters, we fit the measurements in our model and confirm the linear growth of $N_\text{noise}$. 
It is notable that pulsed pump at constant average power presents better CAR than CW regime due to improved SHG generation and thus higher photon pairs per pulse, while the accidentals from $N_\text{noise}$ are slightly lower (see Supplementary). 
Furthermore, stronger conversion efficiencies shift the CAR peak towards lower pump power.

We also quantify the quality of entanglement for a coarse WDM scheme. The broadest filter we can set for our source with our programmable filter is 8 nm bandwidth, although broader channels should also be possible with different DEMUX. 
We use this setting for both the signal and idler, separated by a notch of 1.6 nm (as shown in the light gray area in Fig. \ref{fig:Spectra}). 
The pump power is set to 9.0 dBm just before the PBS at an 8-GHz pulsed regime and 50\% duty cycle, the fastest pump rate we can set with our current electrical devices.  
We observe  visibilities of $V_{HV}=99.4\%$ and $V_{DA}=97.7 \%$, with no subtraction of accidentals. The lower bounded fidelity is thus $F\geq98.6\%$. 

The detected pair rate at 12.85 dB loss is $R_{HH}=$8027 and $R_{VV}=$7812, with a heralding of 1.7\% for the signal and 3.4\% for the idler photon with respect to the averaged $|HH\rangle$ and $|VV\rangle$ projection settings. 
The actual emission rates can be estimated by compensating for the loss such that $R^\text{est}=R^\text{exp}/\eta_s\eta_i$, with $R^\text{exp}$ the experimental measurement, $\eta_s$=6.6\%, $\eta_i$=4.1\% (see Methods), and amounts to 5.9 Mcps at this power and channel bandwidth. %Similarly, we can also estimate the heralding at emission (at the output of the PBS) as $H_s^{est}=H_s^{exp}/\eta_i$ and $H_i^{est}=H_i^{exp}/\eta_s$.
By replacing several parts of our detection system with commercially available low-loss components, the losses could be reduced at least by 10 dB. 
In that case, the usable photon pairs would be in the order of 1.6 Mcps with a heralding of 20-30\%, although such high photon rates would already saturate all current single photon detectors. Typical InGaAs avalanche photodiodes have deadtimes of the order of 10-20 $\mu$s, while SNSPDs deadtimes are around 1 $\mu$s. Therefore, one would need to work with lower power values.
The emitted brightness with respect to the 1560 nm pump is 93.3 kcps/mW². However, to compare our system with other $\chi^{(2) }$ sources, the estimated power of the second harmonic at 780 nm can be used for normalization, which is 3 $\mu$W per SHG module at this pump power. In this case, the emitted total pair rate corresponds to 1 Gcps/mW. To conclude, we summarize the results of this section in Table \ref{table:source_data}.

This characterization shows that brightness and heralding in our source improve with shorter duty cycle and that our design can be used for any repetition rate (as long as fiber dispersion does not affect). 
In our experiment, the highest repetition rate (8 GHz) and shortest pulse width ($\sim$90ps) are driven at the limit of our electrical pulse generator. Thus, deteriorated electrical signals eventually deform the optical pulse, leading to lower spectral brightness. Yet, our design is not only restricted to pulsed operation, since good figures of merit are also obtained for CW pump. 
In that case, cheap laser diodes could be employed instead of our involved pulse preparation scheme (see Methods).
Additionally, since our source spans over 60 nm of bandwidth, multiple demultiplexed channels can be used with large photon numbers and potentially high interference visibility. We note that a better spectral overlap between both ppLN modules would be needed to achieve high visibility.

\begin{figure*}[ht]
\centering
\includegraphics[width=0.8\textwidth]{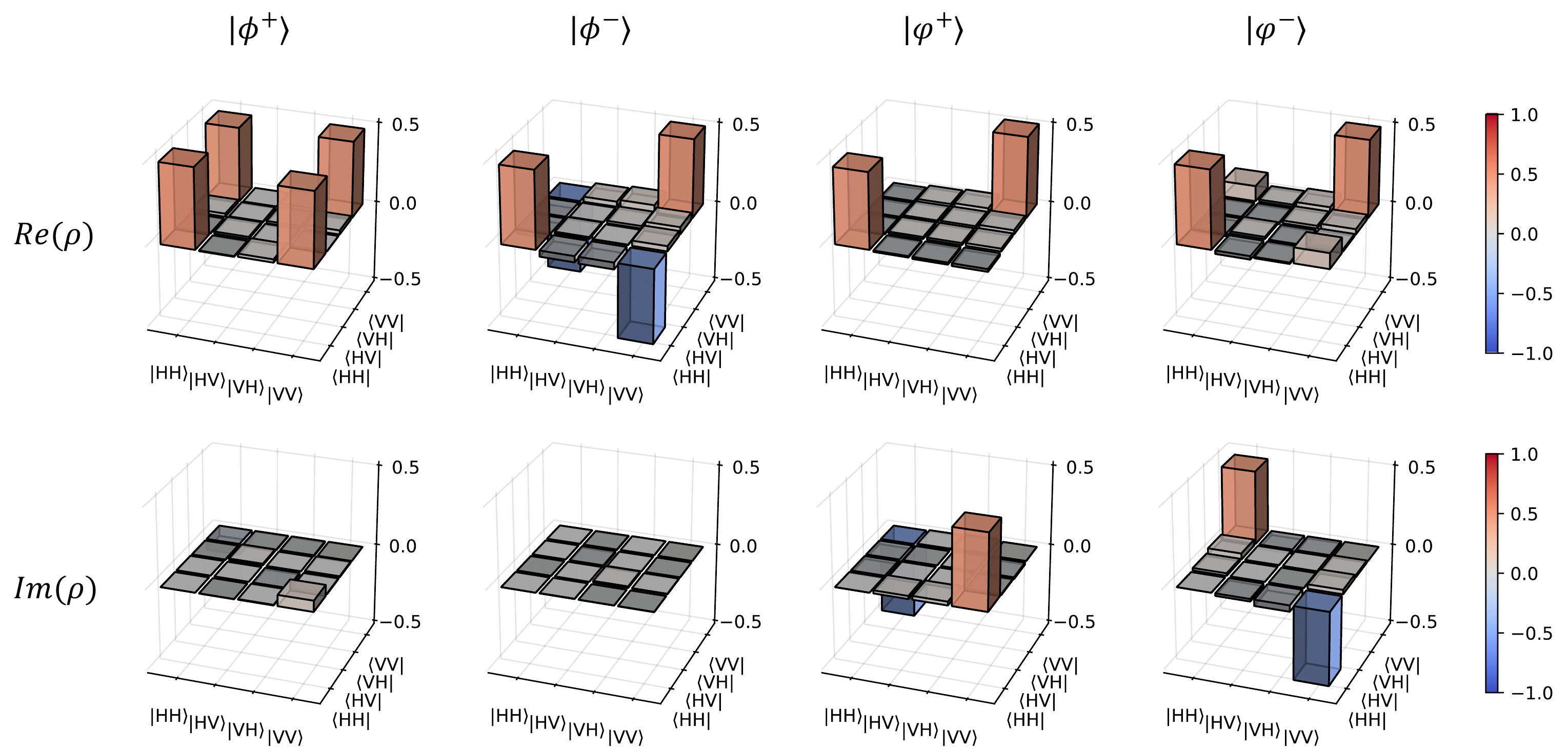}
    \caption[Phase]{Reconstructed density matrices of the four selected states with overall fidelities of 95\%. On the first row, the real part of the density matrix is displayed, while the second raw shows the imaginary part. Notice how the diagonal terms are always the same, while the off-diagonal terms indicate the phase of the state.}
    \label{fig:Phase}
\end{figure*}

\section{Active state control}
Finally, we investigate the fast switching capabilities of the polarization-entangled state phase as a basic prerequisite for active quantum networks. For this purpose, a phase modulator and a PM fiber delay line are inserted between the ppLN waveguides, as shown in Fig. \ref{fig:Setup}. 
In this manner, the SHG pulses that travel along the clockwise direction traverse the EOM first, which introduces a phase $\theta_\text{early}$, followed by the delay line. 
The pulses traveling in the counter-clockwise direction encounter an additional delay before they acquire the phase $\theta_\text{late}$ from the EOM. 
Since the length of the interferometer is the same in both directions, the SPDC probability amplitudes of both paths are perfectly recombined at the output even after SHG modulation. 

The final state depends on the relative phase between the SHG pulses. Therefore, it is enough to modify the phase of only one of them (lower right inset in Fig.\ref{fig:Setup}). When no voltage is supplied to the EOM, and if a diagonal pump polarization is fed into the loop as described previously, the state remains the same:

\begin{equation}
| \phi^+ \rangle = \frac{1}{\sqrt{2}}  \left( |HH\rangle + |VV\rangle \right)
\end{equation}

However, if $V_{\pi}$ is applied to one of the pulses, i.e. $\vartheta = \theta_\text{early} - \theta_\text{late} = \pi$, then the entangled state becomes:

\begin{equation}
| \phi^- \rangle = \frac{1}{\sqrt{2}}  \left( |HH\rangle - |VV\rangle \right)
\end{equation}
which is another maximally entangled Bell state. Nonetheless, we can also generate other states that are necessary for cryptographic primitives such as quantum secret sharing  \cite{williams_quantum_2019}. We choose the “$i+/i-$” states:

\begin{equation}
| \varphi^+ \rangle = \frac{1}{\sqrt{2}}  \left( |HH\rangle + i |VV\rangle \right)
\end{equation}

\begin{equation}
| \varphi^- \rangle = \frac{1}{\sqrt{2}}  \left( |HH\rangle - i |VV\rangle \right)
\end{equation}
that just require $\pm V_{\pi/2}$ voltage for the EOM. If the phase is introduced to the first pulse, the sign of the voltage transfers to the phase in the superposition. If introduced to the second pulse, then the sign of the phase is inverted. One can also restrict only to positive voltages by applying $+ V_{\pi/2}$ either to the first or the second SHG pulse to alternate between these two states.

We perform the phase modulation experiment by using the second channel of our electrical signal generator, which allows us to fine-tune the time and amplitude of the RF electrical pulses that feed the EOM. 
These pulses are generated at a 100 MHz rate and have a duty cycle of $5\%$. 
The pulse amplitude is chosen from the voltage set $\{ 0, +V_\pi, -V_{\pi/2}, +V_{\pi/2}\} $, targeting the second and delayed SHG pulse (see Methods). 
The optical pulses are generated at the same repetition rate and duty cycle of $1\%$, with an average power of 5 dBm before the PBS. Note that the losses of the EOM account for 3.5 dB, but only affect the SHG stage and do not deteriorate the final entangled state. 
The filter settings are again 8 nm bandwidth for the signal and idler channels, as in Fig. \ref{fig:Spectra}, and their polarization state is analyzed in our detection system (see Methods).
The measured coincidence rate reaches 4.5 kcps per polarization setting, again after $\sim$12.85 dB loss in our setup. 
We measure the full-state tomography to reconstruct the density matrix and apply maximum likelihood estimation to obtain a physical one \cite{james_measurement_2001, kwiat_quantum_information_group_quantum-tomography_2021}. The final results are presented in Fig. \ref{fig:Phase}.

With little optimization effort, we achieve fidelities of $95.0\%, 95.0\%, 95,8\%$ and $94.3\%$ for the states $|\phi^+\rangle$, $|\phi^-\rangle$, $|\varphi^+\rangle$ and $|\varphi^-\rangle$ respectively. These measurements are performed in one run by integrating over 100 ms per polarization setting and averaging over 10 samples, while the same periodic RF electrical pulse is employed continuously to modulate the state phase. To change the state, we only change the pulse sequence at the electrical pulse generator.
Nevertheless, our signal generator and common RF sources can also be fed with a very long sequence of arbitrary pulses among the set of four. In that case, a state switching at a 100 MHz repetition rate would occur. Even higher switching rates up to GHz are possible, but are eventually limited to the propagation time of the pulses in the phase modulator. 

\section{Conclusion}
We have developed and experimentally evaluated a fiber-based Sagnac interferometer that utilizes standard telecom components to generate high-quality polarization-entangled states for a variety of pulse rates and duty cycles. 
We show that our source can be easily employed in CW up to GHz-pulsed regimes, with the latter being more suitable due to the stronger SHG and SpRS noise reduction and hence higher brightness and CAR. 
We obtain fidelities of  0.99 and generate millions of photon pairs per nm of spectral bandwidth, while the source spectrum extends 60 nm. 
Although our detection rates are considerably affected by the loss in the optical components used to characterize the source, most notably the DEMUX, these numbers would be substantially improved
using commercially available low-loss components. We also prove great entanglement quality after active state switching at 100 MHz rates with fidelities up to 0.95.  While not experimentally demonstrated here, we note that the source scheme is also ideally suited to the generation of hyperentanglement in the time-energy and polarization degrees of freedom, as well as other hybrid quantum states. To prove this, only the detection scheme should be adapted.

Combined with the simplicity of the design and its compact footprint, these results make the source an ideal device for future field experiments on quantum entanglement tests and quantum communication networks. 
Particularly, the use of standard telecom components is well suited for integration in compact CubeSat platforms for satellite-based quantum networks. 
The versatile spectrally broadband source could be used for multi-user operation with variable temporal and spectral bandwidth allocation or high-capacity backbone channels with frequency division multiplexing, making it an ideal candidate for integration into flexible quantum network architectures of the future.

\section*{Methods}
\textbf{Pump preparation}
A continuous-wave telecom laser operating at a center wavelength of 1560 nm is used as an initial pump source. If optical pulses are required for a certain application, they are carved from the laser by an intensity modulator (IM), controlled by an arbitrary waveform generator (AWG Tektronik 70002B), and amplified by an erbium-doped fiber amplifier (EDFA). 
The IM bias control and the total optical power are monitored by tapping a fraction of the optical power via fiber beam splitters (FBS). 
A tunable optical bandpass filter (BPF) placed thereafter spectrally purifies the pulses by removing the out-of-band amplified spontaneous emission (ASE) originating from the amplifier. It is worth noting that the BPF (0.2 nm bandwidth) placed after the EDFA is narrower than the notch filter after the Sagnac loop to minimize any pump noise coupled to the detection system.

\textbf{Detection system}
A polarization-independent programmable filter (Finisar WaveShaper 16000A) is used to separate the signal and idler photons for the convenience of wavelength tunability. 
To analyze the polarization entanglement of the photon pairs, we use a free-space polarization analysis module (PAM), consisting of a half-wave plate (HWP) and a free-space PBS. 
This setting allows us to have a constant polarization at the output of the PAM, which is more suitable for efficient detection with our superconducting nanowire single-photon detectors (SNSPD). 
The SNSPDs (SingleQuantum) are optimized for C-band operation and have specified detection efficiencies of $\eta\approx 74 \%$. 
Lastly, a time-to-digital converter (TDC; Qutools QuTAG with 1 ps resolution and 4.2 ps RMS time jitter) analyzes the photon-pair coincidences. 

\textbf{Loss characterization}
From the photon pair sources to the detection system, an average total loss of 12.85 dB are estimated, which breaks down as follows: 1 dB for the CIRC, 1.3dB for the notch filter, 6 dB for the programmable optical filter, 3.25 dB insertion loss for the PAM and 1.3 dB detection loss. However, small asymmetries on each path lead to an overall detection efficiency for the signal channel $\eta_s=$-11.8 dB, whereas for the idler channel it is $\eta_i=$-13.9 dB.
These significant losses can be reduced by at least 10 dB with low-loss components, particularly the DEMUX and PAM parts, but are tolerable to demonstrate the source functionality. We note that DEMUX with less than 1 dB loss are already commercially available. 

\textbf{Active state control}
The phase modulator has $V_\pi$ = 4 V at 50 kHz - 2 GHz (iXblue, specs). The peak voltages used for our phase modulator are $\{ 0, +V_\pi, -V_{\pi/2}, +V_{\pi/2}\} $ = $\{0, 250, -112.5, 112.5\}$ mV. The pulses are then amplified with an RF amplifier with $\sim$28 dB of fixed gain. 
To analyse the polarization of the entangled state, we include now in our previous PAM a quarter wave plate before the HWP to access the phase information.

\section*{Funding}
This work was financially supported by the Federal Ministry of Education and Research of Germany (BMBF) through the QuNET initiative.

\begin{acknowledgments}
The presented results were partially acquired using facilities and devices funded by the Free State of Thuringia within the application center Quantum Engineering. M.C.P and C.S. are part of the Max Planck School of Photonics supported by BMBF, Max Planck Society, and Fraunhofer Society.
\end{acknowledgments}

\bibliography{MyBiblio}% Produces the bibliography via BibTeX.

\clearpage

\appendix*

\section{Single-photon noise analysis }
The Sagnac configuration requires the 1560 nm pump to be introduced in the interferometer via the same components used to extract the SPDC. 
This implies that, even over short distances, the pump and the SPDC may co-propagate along the same fibers. 
In this setting, spontaneous four-wave mixing (SFWM) and spontaneous Raman scattering (SpRS) could potentially occur. 
SFWM results as a parametric interaction of two pump photons with a third-order nonlinear media, such as optical fibers, generating two other new photons at neighboring frequencies \cite{caspani_integrated_2017}.
However, this process can be discarded in our experiment since no time-energy correlations were found among the noise photons. 
This observation agrees with the discussions in Refs. \cite{lin_photon-pair_2007,eraerds_quantum_2010}. To quantify the amount of generated SFWM under phase-matching conditions, the product $\gamma P_0 L$ can be used, where $\gamma$ is the nonlinear fiber parameter (typically $\gamma \sim 1 W^{-1} km^{-1}$  for standard fibers), $P_0$ denotes the pump power, and $L$ stands for the fiber length. For any substantial contribution of SFWM, the product should be $\gamma P_0 L>0.1$ \cite{lin_photon-pair_2007}. In our system, $L\sim10$ m and $P_0=10$ mW, which leads to the product value of $\sim$0.0001, indicating minor SFWM contribution.
While SFWM noise is negligible, we do expect a substantial noise contribution of SpRS. 
This scattering process is known to be an inelastic effect, where a pump photon interacting with the material is annihilated, creating a photon at a lower frequency (Stokes) and a phonon to conserve energy and momentum. 
Photons at higher frequencies (anti-Stokes) can also be generated when existing phonons transfer energy to pump photons. 
These scattering events occur in any direction and due to the non-crystalline nature of silica fibers, they extend over a large frequency range (up to 40 THz or 300 nm), surpassing the SPDC signal \cite{agrawal_nonlinear_2013}. 
Theoretical models referred to in \cite{eraerds_quantum_2010} predict a linear growth of SpRS as a function of initial power and fiber length (i.e. restricted only for short fiber distances).

\begin{figure}[ht]
	\centering
\includegraphics[width=0.49\textwidth]{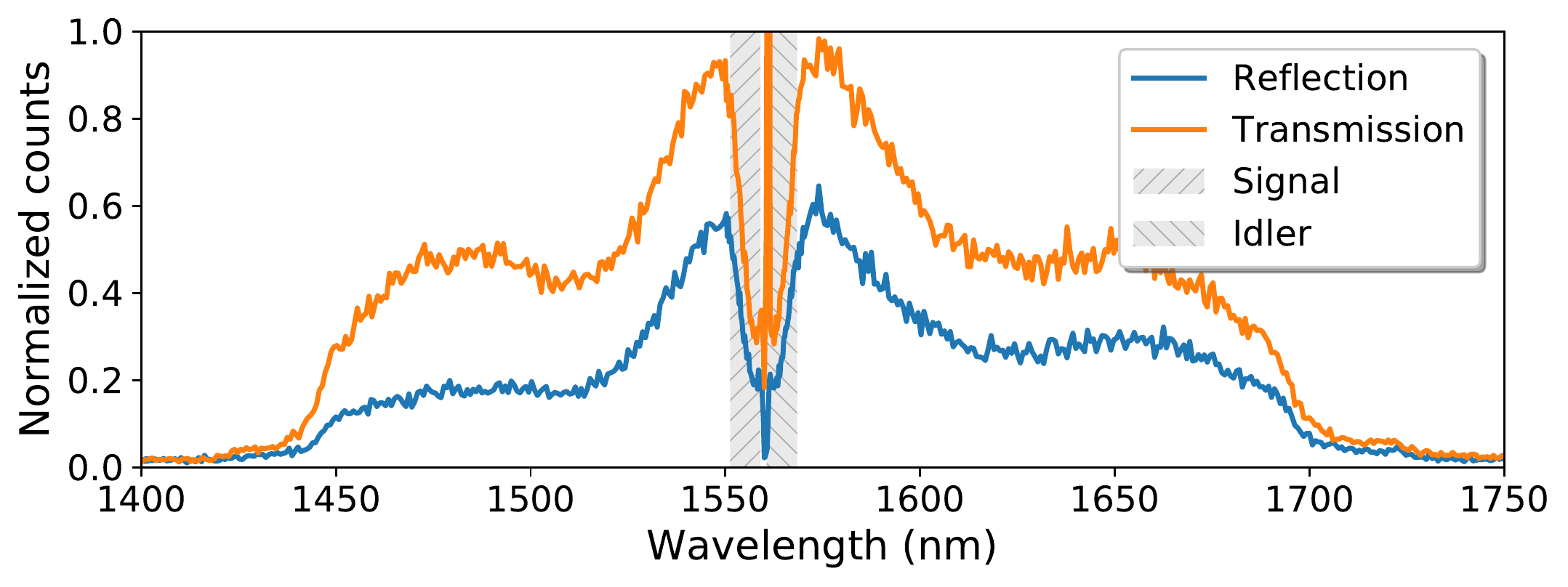}
    \caption{Spectrum of spontaneous Raman scattering after pump propagation through 12m of optical fiber and at room temperature. The 8nm channels of signal and idler from our entanglement characterization are shown in grey for comparison. They coincide with the lowest spectral SpRS contribution.}
    \label{fig:SpRS_spectrum}
\end{figure}

Although SpRS photons are uncorrelated events and do not directly contribute to coincidence detections, a high number of noise photons increases the probability of accidental detections and deteriorate figures of merit, such as CAR and heralding. 
To investigate the behaviour of the noise alone, we isolate it from the SPDC photons by replacing the fiber-loop with a 10 m SMF. 
In this way, we skip the PBS and the up-and down-conversion processes and characterize the noise generated as a function of the initial power and pulse duty cycle. Note that we leave the second fiber end, terminated in APC (angle-polished connector), disconnected and no reflections should be expected from here. 
Fig. \ref{fig:SpRS_spectrum} displays the noise spectrum at room temperature generated over approximately 12 m of accumulated optical fiber from the BPF to the circulator and the 10 m SMF. For the reflection curve (i.e., the second end of the 10 m SMF is left disconnected), a dip in the spectrum at 1560 nm is caused by the spectral filtering of the pump with the notch filter. 
In contrast, for the transmission case (i.e., the second end of the 10m SMF is directly connected to notch filters and to the spectrometer), the pump power is too high and is not completely removed. Similar SpRS spectrum traces have been found in other works \cite{hu_quantum_2020}.

Even though the noise spectral brightness is not very high, the extremely broad bandwidth already hints at its impact on the source. 
In fact, the total number of circulating photons in the system due to SpRS alone is so high that the single-photon detectors are easily saturated. 

\begin{figure}[ht]
    \centering
    \includegraphics[width=0.49\textwidth]{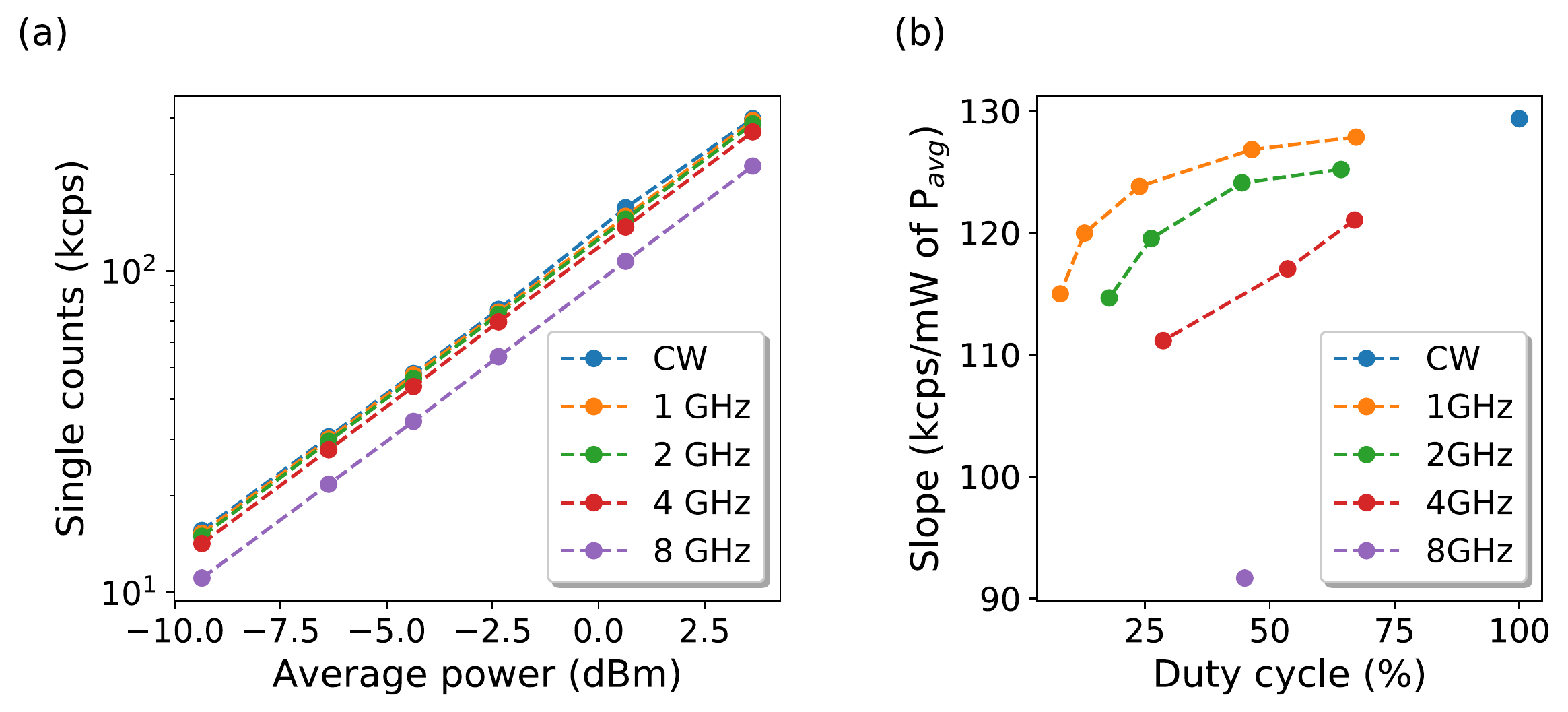}
    \caption{Analysis of SpRS behaviour for different pump configurations: (a) Single-photon detection of SpRS only (i.e. the EPS has been disconnected and replaced with a 10 m fiber with open end). The increase is linear with pump power.}
    \label{fig:SpRS_growth}
\end{figure}

To uncover the main noise trends while avoiding saturation, we introduce now a variable optical attenuator before the SNSPD in the same setup as before, the 10 m SMF instead of the fiber loop, and we measure the single-photon noise in a reflection configuration. 
For clear visualization, we show the noise-photon generation as a function of repetition rate (50 \% duty cycle), and average pump power in Fig. \ref{fig:SpRS_growth}(a), and verify the linear increase with power. In Fig. \ref{fig:SpRS_growth}(b), we summarize and normalize the obtained generation rates for different pulse rates and duty cycle. 
Note that a constant average pump power ensures the same number of pump photons per second, even when the pulse settings are different. Given that SpRS is a stochastic process, the number of scattering events should be proportional to the number of pump photons and thus, it should also be constant. Interestingly, we found that the CW regime presents the highest noise rate, suggesting that the probability of generating SpRS photons is higher than for pulsed signals. Instead, pulsed signals with decreasing duty cycles reveal a slight reduction of the noise, e.g. the 8-GHz pulsed regime appears to be the best scenario with the lowest noise contribution.

In sum, we can conclude that to diminish the impact of SpRS noise, which grows linearly with power, a sufficient strong pump should be used to improve the SHG and thus SPDC processes, which grow quadratically. As already shown in the main article, this regime can be found before generating multiphoton states or stimulated processes that would decrease our visibility measurements. Nonetheless, and perhaps more importantly, spectral filtering is required. For example, by spectrally filtering the 60 nm broad SPDC band, the remaining 200 nm of almost uniformly distributed SpRS noise can be easily suppressed. This procedure can be further optimized if narrower channels are used for filtering, which is perfectly suitable for dense wavelength division multiplexing (DWDM) schemes.

\end{document}